\documentstyle[a4,
twoside]{article}
\makeatletter
\renewcommand{\@oddhead}{ Notes on Rank-One Perturbed Resolvent 
 \hfill \thepage}
\renewcommand{\@evenhead}{\thepage \hfill S. A. Choro\v{s}avin }
\renewcommand{\@oddfoot}{}
\renewcommand{\@evenfoot}{}
\makeatother

\author{S.A.~Choro\v{s}avin}
\title{ 
 Notes on Rank-One Perturbed Resolvent.
 }
\date{}
\begin{document}
\maketitle 
\begin{abstract}
 This paper is a didactic comment (a transcription with variations) 
 to the paper of S.R. Foguel 
 {\it Finite Dimensional Perturbations in Banach Spaces}. 

 Addressed, mainly: postgraduates and related readers.

 Subject:
 Suppose we have two linear operators,
$T_1, T_2$,
 so that 
$$ T_2^{-1} - T_1^{-1} \mbox{ is rank-one. }$$
 What we want to know is 
 how we can express 
$$
 (z-T_2)^{-1}
$$
 in terms of 
$T_2^{-1} - T_1^{-1}$
 and 
$(z-T_1)^{-1}$
 .

\end{abstract}

\newpage
\section*%
{ Introduction }

 The problem I have is that my students can prove theorems, but not invent 
 good formulae, the formulae that are worth to discuss. 
 They, the students, can correct Leibniz, and are not capable 
 of realizing 
 standard and simple engineering computation 
 of primitive engineering problem. 
 They can {\it explain} and {\it teach} what 
 Green's function is indeed and really, 
 and are not able to {\it compute} THE Green's function. 

\addvspace{\medskipamount}\par\noindent
 Perhaps, I am wrong.
 I would be glad, if I were wrong.
 I hope I am wrong.
 However, I have expounded the situation I have observed it.

\addvspace{\medskipamount}\par\noindent
 I cannot solve the problem. 
 But as a contribution I try to tell about some old tricks 
 which can help to {\it compute},
 not to {\it substantiate the being computed and ready}.

 But first of all, let us state a {\it primitive} problem.

\addvspace{\medskipamount}\par\noindent
 Let
$T$
 stands for the functions transformation defined by 
$$
  (Tu)(x) := -\frac{\partial^2 u(x)}{\partial x^2 } \,;
$$
$T_{DD}$
 and 
$T_{DN}$
 be the restrictions of 
$T$
 so that 
$T_{DD}$
 and 
$T_{DN}$
 act on that functions,
$u$,
 which are defined on
$[0,1]$
 and, in addition: 
\begin{eqnarray*}
 u(0) &=& 0
\\
 u(1) &=& 0
\\&&{} \mbox{ in the case of } T_{DD} \,,
\end{eqnarray*}
\begin{eqnarray*}
 u(0) &=& 0
\\
 \frac{\partial u(x)}{\partial x}\Big|_{x=1}&=& 0
\\&&{} \mbox{ in the case of } T_{DN} \,.
\end{eqnarray*}
\footnote{
 take the underlying space which you like, within reason, 
 then make corrections
}
 These
$ T_{DN}^{-1} \,,\, T_{DD}^{-1} $
 are integral operators and their integral kernels are 
\begin{eqnarray*}
 G_{DD}(x,\xi)
 &=&
 -
   \left\{\begin{array}{rcl}
   x\cdot (\xi-1) &,& \mbox{ if }   x \leq \xi \\
   (x-1)\cdot \xi &,& \mbox{ if } \xi \leq x \\
   \end{array}\right\}
\end{eqnarray*}
\begin{eqnarray*}
 G_{DN}(x,\xi)
 &=&
   \left\{\begin{array}{rcl}
     x &,& \mbox{ if }   x \leq \xi \\
   \xi &,& \mbox{ if } \xi \leq x \\
   \end{array}\right\}
\end{eqnarray*}
 Let us draw the attention on that fact that 
$$
 T_{DN}^{-1} - T_{DD}^{-1}
$$ 
 is well defined and its integral kernel is 
\begin{eqnarray*}
 G_{DN}(x,\xi) - G_{DD}(x,\xi) 
 &=&
   \left\{\begin{array}{rcl}
     x + x\cdot (\xi-1) &,& \mbox{ if }   x \leq \xi \\
   \xi +(x-1)\cdot \xi &,& \mbox{ if } \xi \leq x \\
   \end{array}\right\}
 \\&&{} 
 =
   \left\{\begin{array}{rcl}
     x\cdot \xi &,& \mbox{ if }   x \leq \xi \\
     x\cdot \xi &,& \mbox{ if } \xi \leq x \\
   \end{array}\right\}
 = x\cdot \xi
\end{eqnarray*}
 We formulate this result a little generalizing:
\begin{eqnarray*}
 G_{DN}(x,\xi) - G_{DD}(x,\xi) 
 &=&
 f(x)l(\xi)
 \\&&{} 
 \mbox{ where } f(x) = x \,,\, l(\xi) = \xi \,.
\end{eqnarray*}
 So, we state:
\begin{eqnarray*}
 T_{DN}^{-1} - T_{DD}^{-1}
 &=&
 \mbox{ is {\bf rank-one .} }
\end{eqnarray*}
 It is not very difficult to describe 
$$
 (z - T_{DD})^{-1} \,.
$$ 
 This is an integral operator. Its integral kernel is 
\begin{eqnarray*}
 G_{DD}(x,\xi,z)
 &=&
 -\frac{1}{k\sin(k)}
   \left\{\begin{array}{rcl}
   \sin(kx)\sin(k(1-\xi)) &,& \mbox{ if }   x \leq \xi \\
   \sin(k\xi)\sin(k(1-x))  &,& \mbox{ if } \xi \leq x \\
   \end{array}\right\}
 \\&&{} 
 \mbox{ where $k$ is defined by } k^2 = z \,,
\end{eqnarray*}
 and where, of course, 
$z$
 is to be so, that 
$$
 \sin(k) \not= 0 \,.
$$
 As for  
$$
 (z - T_{DN})^{-1} \,,
$$ 
 it is not very difficult to describe it as well, 
 but the primitive problem 
 in this {\it toy} situation 
 is: 

\addvspace{\medskipamount}\par\noindent
{\it
 to reduce {\em(effectively)} the description of 
\mbox{ $ (z - T_{DN})^{-1} $ }
 to the description of 
\mbox{ $ (z - T_{DD})^{-1} $ }
} ;

\addvspace{\medskipamount}\par\noindent
{\it 
 to construct an effective abstraction.
}

\addvspace{\medskipamount}\par\noindent

 A beautiful example of such an abstraction is that what is presented in 
 the paper of 
 {\sc S.R. Foguel}, 
 {\it Finite Dimensional Perturbations in Banach Spaces}.  
 We will not discuss all the contents of that paper 
 and restrict ourselves, centring on its initial part.
 Namely,
 our section 1 is a transcription
 of the section 1 of the {\sc Foguel}'s paper,
 and our section 2 is a variation, or modification,
 adapted to the described situation.

\addvspace{2\bigskipamount}\par\noindent

 Before starting,
 we shall say a few words about features of our paper. 
 The features of the notations we use are: we prefer Dirac's 
 "bra-ket" style of expressing, in the following form:

\addvspace{\bigskipamount}\par\noindent
{\it Notation 1}. \quad 
 If 
$f$
 is an element of a linear space, 
$X$,
 over a field, 
$K$, 
 then 
$|f>$
 stands for the mapping
$K \to X$,
 defined by 
$$
 |f>\lambda := \lambda f \quad .
$$

\addvspace{\bigskipamount}\par\noindent
{\it Notation 2}. \quad 
 If 
$l$
 is a functional 
 and we wish to emphasise this factor, then we write 
$<l|$ 
 instead of 
$l$.
 We also write 
$<l|f>$
 instead of 
$<l||f>1$,
 and write the terms 
$|f><l|$ 
 and 
$f<l|$
 interchangeably: 
$$
 <l|f> \equiv <l||f>1 \equiv l(f) \quad , \quad f<l| \equiv |f><l| \quad .
$$

\addvspace{\bigskipamount}\par\noindent
 Finally, a feature of the paper is that 
 we emphasise the algebraic aspect of the problem 
 and try to deemphasise the topological one.
 We have in mind differetial and integral operators, indeed,
 but we avoid metioning this fact,
 in order to place in the centre algebra, FORMULA.

\newpage 
\section
{\bf Rank-One Perturbations. Abstract Formulae. }
 The situation we will discuss in this section is this.
 Suppose, we deal with two linear operators,
$A$ 
 and 
$B$,
 so that 
$A-B$
 is rank-one
\footnote{ more accurately expressed, rank-one or less },
 i.e.,
$$
 B-A = -f_a<l_a|
$$
 for an element 
$f_a$
 and a linear functional 
$l_a$.
 The first question is: If 
$A^{-1}$
 exists and is given, what are the conditions
 in order that
$B^{-1}$
 should exist?
 And if
$B^{-1}$
 exists, how can we calculate 
$$
 B^{-1}-A^{-1} \qquad ?
$$
 A possible answer can be found in the following way:

\addvspace{2\bigskipamount}

$B$
 is defined by solving the equation 
$$
 Bv=w
$$
 with respect to 
$v$.
 That is, in the described situation, 
$B$
 is defined by solving the equation 
$$
 Av -f_a<l_a|v>=w
$$
 Firstly we write this equation as 
$$
 Av =w+f_a<l_a|v>
$$
 and then, 
 --recall that 
$A^{-1}$
 is given,-- as 
$$
 v =A^{-1}(w+f_a<l_a|v>)
$$
 Thus we observe: in order to find 
$v$,
 it is sufficient to find
$$
 c_a := <l_a|v> \,;
$$
 we emphasise it:
$$
 v =A^{-1}(w+c_a f_a) \,.
$$
 And if we have found 
$v$,
 then we can found 
$c_a$:
$$
 c_a = <l_a|v> = <l_a|A^{-1}(w+c_a f_a)>
$$
 So, we have obtained the equation to 
$c_a$,
 and we are solving it.

 It is not difficult, because 
$A^{-1}$
 and 
$l_a$
 are linear. Using this factor we deduce:

$$
 c_a = <l_a|v> = <l_a|A^{-1}w+c_aA^{-1}f_a>
$$

$$
 c_a = <l_a|v> = <l_a|A^{-1}w>+c_a<l_a|A^{-1}f_a>
$$

$$
 (1-<l_a|A^{-1}f_a>)c_a = <l_a|A^{-1}w>
$$

\noindent
 Thus, we conclude that 

$$
 c_a = \frac{<l_a|A^{-1}w>}{1-<l_a|A^{-1}f_a>}
 \mbox{ , if $1-<l_a|A^{-1}f_a> \not= 0 $}
$$
 and then 
$$
 v = A^{-1}w+A^{-1}f_a\frac{<l_a|A^{-1}w>}{1-<l_a|A^{-1}f_a>}
 \mbox{ , if $ 1-<l_a|A^{-1}f_a> \not= 0 $} \,.
$$

\noindent
 We see now that 

\medskip\noindent
\fbox{
\parbox{\textwidth}{
$$
 B^{-1}-A^{-1} = \frac{A^{-1}f_a<l_a|A^{-1}}{1-<l_a|A^{-1}f_a>}
 \mbox{ , if $ 1-<l_a|A^{-1}f_a> \not= 0 $} \,.
$$ 
} 
} 

\addvspace{2\bigskipamount}

 So, we have partially answered the question 
 we have formulated at the beginning of this section.
 Namely, we have found an answer to the question in that case where 
$ 1-<l_a|A^{-1}f_a> \not= 0 $ .

\addvspace{2\bigskipamount}

\noindent
 And what can we state, if 
$$
 1-<l_a|A^{-1}f_a> = 0 \qquad ?
$$
 In this case,
$$
 ( A -f_a<l_a| )A^{-1}f_a = A A^{-1}f_a - f_a<l_a|A^{-1}f_a> =f_a-f_a = 0
$$
 i.e.,

\medskip\noindent
\fbox{
\parbox{\textwidth}{
$$
 \mbox{ if }
 1-<l_a|A^{-1}f_a> = 0
 \mbox{ then }
 B A^{-1}f_a =0 
$$
} 
} 

\medskip\noindent
 An interesting detail is: Let 
$$
 B v_0 =0 
$$
 It means that 
$$
 Av_0 -f_a<l_a|v_0> =0
$$
 and it implies that 
$$
 v_0 = A^{-1}f_a<l_a|v_0>
$$
 Hence 
$$
 <l_a|v_0> = <l_a|A^{-1}f_a><l_a|v_0>
$$
 and
$$
 0 = Bv_0 \equiv B A^{-1}f_a<l_a|v_0> =0 \,.
$$
 Thus we infer:

\medskip\noindent
\fbox{
\parbox{\textwidth}{
 \mbox{ if }
$$
 Bv_0 = 0
 \mbox{ and }
 v_0 \not= 0
$$
 \mbox{ then }
$$
 <l_a|v_0>\not=0 \,,\, 1-<l_a|A^{-1}f_a> =0 \,,\, BA^{-1}f_a =0\,,
 \mbox{ and }
 v_0 = A^{-1}f_a<l_a|v_0> \,.
$$
} 
} 

\addvspace{2\bigskipamount}

 So, we have completely answered the question 
 we have formulated at the beginning of this section.

\newpage\section
{\bf Rank-One Perturbed Resolvent. }
 Suppose we have two operators,
$T_1, T_2$,
 so that 
$$
 T_2^{-1} - T_1^{-1} = |f><l| \,.
$$
 What we want to know is 
 how we can express 
$$
 (z-T_2)^{-1} - (z-T_1)^{-1}
$$
 in terms of 
$T_2^{-1} - T_1^{-1}$
 and 
$(z-T_1)^{-1}$
 .

 We begin the analysis with a general (and quite standard) argumentation:
\begin{eqnarray*}
\makebox[5ex][l]{$\displaystyle (z-T_2)^{-1} - (z-T_1)^{-1} $}
\\
&=&
 T_2^{-1}(zT_2^{-1}-I)^{-1} - T_1^{-1}(zT_1^{-1}-I)^{-1}
\\&&{}=
\frac{1}{z}\Bigl(
 zT_2^{-1}(zT_2^{-1}-I)^{-1} - zT_1^{-1}(zT_1^{-1}-I)^{-1}
\Bigr)
\\&&{}=
\frac{1}{z}\Bigl(
 (zT_2^{-1}-I+I)(zT_2^{-1}-I)^{-1} - (zT_1^{-1}-I+I)(zT_1^{-1}-I)^{-1}
\Bigr)
\\&&{}=
\frac{1}{z}\Bigl(
 (zT_2^{-1}-I)^{-1} - (zT_1^{-1}-I)^{-1}
\Bigr) \,.
\end{eqnarray*}
 Notice now, that 
\begin{eqnarray*}
 \Bigl(zT_2^{-1}-I\Bigr) -\Bigl(zT_1^{-1}-I\Bigr) &=& - (-z)|f><l|
\end{eqnarray*}
\begin{eqnarray*}
 \Bigl(zT_2^{-1}-I\Bigr) &=& \Bigl(zT_1^{-1}-I\Bigr) - (-z)|f><l| \quad .
\end{eqnarray*}
 Now recall the relation, which we have seen in the previous section:
$$
\fbox{ 
$\displaystyle
 (A-f_a<l_a|)^{-1} - A^{-1}
  = \frac{A^{-1}f_a<l_a|A^{-1}}{ 1-<l_a|A^{-1}f_a> }\,, 
 \mbox{ if } 1-<l_a|A^{-1}f_a> \not= 0
$}
$$
 Thus, letting us put 
$$
 A := zT_1^{-1}-I \,,
$$
 we infer:
\begin{eqnarray*}
\makebox[5ex][l]{$\displaystyle (z-T_2)^{-1} - (z-T_1)^{-1} $}
\\
&=&
\frac{1}{z}
\Bigl(
 (zT_2^{-1}-I)^{-1} - (zT_1^{-1}-I)^{-1}
\Bigr)
\\&&{}=
\frac{1}{z}
 \frac{
 (zT_1^{-1}-I)^{-1}|-zf><l|(zT_1^{-1}-I)^{-1}
      }{ 1-<l|(zT_1^{-1}-I)^{-1}(-zf)>}
\\&&{}=
 -\frac{
 (zT_1^{-1}-I)^{-1}|f><l|(zT_1^{-1}-I)^{-1}
      }{ 1+z<l|(zT_1^{-1}-I)^{-1}f>}
\\&&{}=
 -\frac{
 T_1(z-T_1)^{-1}|f><l|T_1(z-T_1)^{-1}
      }{ 1+z<l|T_1(z-T_1)^{-1}f>}
\\&&{}=
 -\frac{
 \bigl(-I+z(z-T_1)^{-1}\bigr)|f><l|\bigl(-I+z(z-T_1)^{-1}\bigr)
      }{ 1+z<l|\bigl(-I+z(z-T_1)^{-1}\bigr)f>}
\\&&{}
 \mbox{ if } 1+z<l|\bigl(-I+z(z-T_1)^{-1}\bigr)f> \not= 0 \,. 
\end{eqnarray*}

\bigskip

 A result is: 

\medskip\noindent
\fbox{
\parbox{\textwidth}{
\begin{eqnarray*}
 (z-T_2)^{-1} - (z-T_1)^{-1}
&=&
 -\frac{
 \bigl(-I+z(z-T_1)^{-1}\bigr)|f><l|\bigl(-I+z(z-T_1)^{-1}\bigr)
      }{ 1+z<l|\bigl(-I+z(z-T_1)^{-1}\bigr)f>}
\\&&{}
 \mbox{ if } 1+z<l|\bigl(-I+z(z-T_1)^{-1}\bigr)f> \not= 0 \,. 
\end{eqnarray*}
} 
} 

\addvspace{2\bigskipamount}

 Now suppose, that although we know that 
$$
 T_2^{-1} - T_1^{-1} = |f><l| \,,
$$
 however
 we do not separately know 
$f$
 or
$l$.
 Let us try to find a method to calculate the value of the expression 
$$
 <l|\bigl(-I+z(z-T_1)^{-1}\bigr)f>
$$
 without 
 specifying 
$f$
 and 
$l$
 separately.
 A way is this. Let
$f_0$
 be an element, {\it arbitrary} taken from the domain of 
$ T_2^{-1} - T_1^{-1} \equiv |f><l| $,
 and 
$l_0$
 be a {\it linear} functional,
 which is defined on the range of 
$ T_2^{-1} - T_1^{-1} \equiv |f><l| $.
 In other words, 
$f_0$
 is an element of the domain of 
$l$,
 and 
$l_0$
 is defined at 
$f$.

 Then we infer: 
$$
 T_2^{-1} - T_1^{-1} = |f><l| \,,
$$
$$
 \Bigl(T_2^{-1} - T_1^{-1}\Bigr)f_0 = |f><l|f_0> \,,
$$
$$
 <l_0|\Bigl(T_2^{-1} - T_1^{-1}\Bigr) = <l_0|f><l| \,,
$$
$$
 <l_0|\Bigl(T_2^{-1} - T_1^{-1}\Bigr)f_0> = <l_0|f><l|f_0> \,.
$$
 Hence, 
\begin{eqnarray*}
 \Bigl(T_2^{-1} - T_1^{-1}\Bigr)f_0
 <l_0|\Bigl(T_2^{-1} - T_1^{-1}\Bigr)
&=&
 |f><l|f_0><l_0|f><l|
\\
&=&
 |f><l_0|\Bigl(T_2^{-1} - T_1^{-1}\Bigr)f_0><l|
\\
&=&
 <l_0|\Bigl(T_2^{-1} - T_1^{-1}\Bigr)f_0>|f><l|
\\
&=&
 <l_0|\Bigl(T_2^{-1} - T_1^{-1}\Bigr)f_0>\Bigl(T_2^{-1} - T_1^{-1}\Bigr) \,.
\end{eqnarray*}
 We have now seen: if we are able to find 
$f_0$
 and 
$l_0$
 so that 
$$
 <l_0|\Bigl(T_2^{-1} - T_1^{-1}\Bigr)f_0> \not =0 \,, 
$$
 then 
$$
 T_2^{-1} - T_1^{-1} = |f_1><l_1| \,,
$$
 where 
$f_1$
 and 
$g_1$
 are defined, e.g., by:
\begin{eqnarray*}
 <l_1|
&:=&
 <l_0|\Bigl(T_2^{-1} - T_1^{-1}\Bigr) \,,
\\
 f_1
&:=&
 \frac{\Bigl(T_2^{-1} - T_1^{-1}\Bigr)f_0}
{<l_0|\Bigl(T_2^{-1} - T_1^{-1}\Bigr)f_0>}
\quad .
\end{eqnarray*}
 Thus we have actually no need to know {\it separetely} 
$f$
 or 
$l$.

\addvspace{2\bigskipamount}

 Another way to calculate the value of the expression of the form 
$ <l|Sf> $
 is:
\begin{eqnarray*}
 <l_0|\Bigl(T_2^{-1} - T_1^{-1}\Bigr)S\Bigl(T_2^{-1} - T_1^{-1}\Bigr)f_0>
&=&
 <l_0|f><l|Sf><l|f_0> \,,
\\
\end{eqnarray*}
\begin{eqnarray*}
 <l_0|\Bigl(T_2^{-1} - T_1^{-1}\Bigr)f_0>
&=&
 <l_0|f><l|f_0> \,,
\\
\end{eqnarray*}
\begin{eqnarray*}
 <l|Sf>
&=&
 \frac%
{<l_0|\Bigl(T_2^{-1} - T_1^{-1}\Bigr)S\Bigl(T_2^{-1} - T_1^{-1}\Bigr)f_0>}%
{<l_0|\Bigl(T_2^{-1} - T_1^{-1}\Bigr)f_0>}
 \quad .
\\
\end{eqnarray*}
 Of course, we have here assumed that 
$$
 <l_0|\Bigl(T_2^{-1} - T_1^{-1}\Bigr)f_0> \not= 0 \,.
$$

\bigskip
 Finally, notice that the formula of
$$
 (z-T_2)^{-1} - (z-T_1)^{-1}
$$
 can be written as:
\begin{eqnarray*}
 (z-T_2)^{-1} - (z-T_1)^{-1}
&=&
 -\frac{
 \bigl(-I+z(z-T_1)^{-1}\bigr)
 \bigl(T_2^{-1} - T_1^{-1}\bigr)
 \bigl(-I+z(z-T_1)^{-1}\bigr)
      }{ 1+z<l|\bigl(-I+z(z-T_1)^{-1}\bigr)f>}
\\&&{}
 \mbox{ if } 1+z<l|\bigl(-I+z(z-T_1)^{-1}\bigr)f> \not= 0 \,. 
\end{eqnarray*}

\newpage\section
{\qquad $ (z - T_{DN})^{-1} - (z - T_{DD})^{-1} $ .
 \mbox{( Only Computations )}
}
 Let 
$$
 f_z := z(z-T_{DD})^{-1}f \,,
$$
 i.e.,
\begin{eqnarray*}
 zf_z(x) + \frac{\partial^2 f_z(x)}{\partial x^2}
&=&
 zx \,,
\\
 f_z(0) 
&=&
 0 \,,
\\
 f_z(1) 
&=&
 0 \,.
\end{eqnarray*}
 Hence 
\begin{eqnarray*}
 f_z(x)
&=&
 x - \frac{\sin(kx)}{\sin(k)}
\\&&{}
 \mbox{ where $k$ is defined by } k^2 = z \,,
\end{eqnarray*}
 and where, recall, 
$z$
 is such that 
$$
 \sin(k) \not= 0 \,.
$$
 Thus we deduce:
\begin{eqnarray*}
 \Bigl((-I+z(z-T_{DD})^{-1})f\Bigr)(x)
&=&
 -f(x) +f_z(x)
\\
&=&
 -x + \Bigl(x -\frac{\sin(kx)}{\sin(k)}\Bigr)
\\
&=&
 -\frac{\sin(kx)}{\sin(k)}
\\&&{}
 \mbox{ where $k$ is defined by } k^2 = z \,,
\end{eqnarray*}
\begin{eqnarray*}
 <l|(-I+z(z-T_{DD})^{-1})f>
&=&
 -\int_0^1 \xi \frac{\sin(k\xi)}{\sin(k)} d\xi
\\
&=&
 \int_{\xi=0}^1 \xi \frac{d\cos(k\xi)}{k\sin(k)}
\\
&=&
 \frac{\cos(k)}{k\sin(k)}-\int_{\xi=0}^1 \frac{\cos(k\xi)}{k\sin(k)}d\xi
\\
&=&
 \frac{\cos(k)}{k\sin(k)}- \frac{1}{k^2}
\end{eqnarray*}
\begin{eqnarray*}
 1+z<l|(-I+z(z-T_{DD})^{-1})f>
&=&
 1+k^2\Bigl(\frac{\cos(k)}{k\sin(k)}- \frac{1}{k^2}\Bigr)
\\
&=&
 k\frac{\cos(k)}{\sin(k)}
\end{eqnarray*}

\addvspace{\medskipamount}\par\noindent

 {\it We conclude} :

\addvspace{\medskipamount}\par\noindent
 The new eigenvalues,
$z_n$,
 are defined by 
$$
 1+z_n<l|(-I+z_n(z_n-T_{DD})^{-1})f> =0 \,,
$$
 i.e., by 
$$
 \cos(k_n) =0 \,,
$$
 and the associated eigenfunctions are 
\begin{eqnarray*}
 \Bigl((-I+z_n(z_n-T_{DD})^{-1})f\Bigr)(x)
&=&
 -\frac{\sin(k_n x)}{\sin(k_n)} \,.
\end{eqnarray*}
 Finally,
 the integral kernel of 
$$
 (z - T_{DN})^{-1} - (z - T_{DD})^{-1}
$$
 is
\begin{eqnarray*}
 G_{DN}(x,\xi,z) - G_{DD}(x,\xi,z)
&=&
 -\frac{\sin(kx)\sin(k\xi)}{k\sin(k)\cos(k)}
\end{eqnarray*}

\bibliographystyle{unsrt}

\end{document}